\documentclass[sigplan,screen,authorversion,nonacm,pbalance]{acmart}
\usepackage{fancyhdr}
\AtBeginDocument{%
    \addtolength{\footskip}{2.0\baselineskip}%
    \fancyfoot[L]{\textit{Preprint}}%
}

\usepackage[utf8]{inputenc}
\usepackage[T1]{fontenc}

\usepackage{xspace}

\newcommand{\dia}[1]{\ensuremath{\langle#1\rangle}}
\newcommand{\backslashLBracket}{\ensuremath{[}}
\newcommand{\backslashRBracket}{\ensuremath{]}}
\newcommand{\dlf}[2]{\langle\java{#1}\rangle{#2}}
\newcommand{\dlfunc}[1]{\mathit{#1}}
\newcommand{\dlbox}[1]{\ensuremath{\backslashLBracket#1\backslashRBracket}}
\def\cdlbr{\hbox{\texttt{\char`\{}}}
\def\cdrbr{\hbox{\texttt{\char`\}}}}
\newcommand{\cd}[1]{\text{\let\{\cdlbr\let\}\cdrbr
  \normalfont\mdseries\ttfamily\upshape#1}}
\newcommand{\upd}{\mathrel{:=}}
\newcommand{\upl}{\{}
\newcommand{\upr}{\}}
\newcommand{\keyeq}{\ensuremath{\doteq}}
\newcommand{\keytrue}{\mathop{\mathrm{true}}}

\newcommand{\keyboolfalse}{\mathit{FALSE}}
\newcommand{\keybooltrue}{\mathit{TRUE}}
\newcommand{\turnstyle}{\Longrightarrow}
\newcommand{\ruleName}[1]{{\ensuremath{\mathsf{#1}}}}
\newcommand{\seq}[2]{#1\turnstyle#2}

\newcommand{\sequent}[2]{\seq{\Gamma\ifx#1\relax\else,#1\fi} 
                             {\ifx#2\relax\else#2,\fi\Delta}}

\newcommand{\adfrac}[2]{\genfrac{}{}{}{0}
  {\begin{array}{l}#1\end{array}}
  {\begin{array}{l}#2\end{array}}}

\newcommand{\seqRule}[3]{\mbox{\small\ruleName{#1}}\ 
  \adfrac{#2}{#3}}
\newcommand{\java}[1]{\cd{#1}}

\newcommand{\KeY}{\texorpdfstring{Ke\kern-0.1emY}{KeY}\xspace}
\newcommand{\parupd}{\,\|\,}
\newcommand{\post}{\mathit{post}}
\newcommand{\pre}{\mathit{pre}}

\usepackage{listings}
\makeatletter
\lstdefinelanguage[JML]{Java}{%
        alsoletter = {\\},%
        morekeywords= [1]{%
                        break,case,catch,class,%
                        const,continue,default,do,else,extends,false,%
                        finally,for,goto,if,implements,import,instanceof,%
                        interface,new,null,package,
                        return,super,switch,this,throw,%
                        throws,true,try,void,while},%
        morekeywords = [2]{%
                        boolean,byte,char,double,float,int,long,short,\\bigint,\\locset,\\real,\\seq,\\TYPE},%
        morekeywords = [3]{%
                        abstract,code,code_bigint_math,code_java_math,code_safe_math,
                        extract,final,ghost,helper,instance,model,native,non_null,nullable,nullable_by_default,private,protected,
                        peer,\\peer,public,pure,rep,\\rep,spec_bigint_math,spec_java_math,spec_protected,
                        spec_public,spec_safe_math,static,strictfp,strictly_pure,synchronized,
                        two_state,
                        uninitialized,volatile,
                        no_state,modifies,erases,modifiable,returns,break_behavior,continue_behavior,return_behavior},%
        morekeywords = [4]{%
                        \\constraint_for,\\created,\\disjoint,\\duration,\\everything,\\exception,\\exists,\\forall,\\fresh,
                        \\index,\\invariant_for,
                        \\is_initialized,\\itself,
                        \\lblneg,\\lblpos,\\lockset,\\max,\\measured_by,\\min,\\new_elems_fresh,\\nonnullelements,
                        \\not_accessed,\\not_assigned,\\not_modified,\\not_specified,\\nothing,\\num_of,
                        \\old,\\only_assigned,\\only_called,\\only_captured,\\pre,\\product,\\reach,\\reachLocs,\\result,
                        \\same,\\seq_contains,\\space,\\static_constraint_for,\\static_invariant_for,\\strictly_nothing,\\subset,\\sum,\\type,\\typeof,\\working_space,\\values,\\array2seq,%
                        \\at,\\backup,\\perm,\\readPerm,\\transactionUpdated,\\writePerm%
                },%
        morekeywords = [5]{%
                        accessible,accessible_redundantly,assert,assert_redundantly,assignable,
                        assignable_redundantly,assume,assume_redudantly,breaks,breaks_redundantly,
                        \\by,
                        callable,callable_redundantly,captures,captures_redundantly,continues,continues_redundantly,
                        debug,\\declassifies,
                        decreases,decreases_redundantly,decreasing,decreasing_redundantly,diverges,
                        determines,
                        diverges_redundantly,duration,duration_redundantly,ensures,
                        ensures_redundantly,\\erases,forall,
                        for_example,hence_by,implies_that,in,in_redundantly,\\into,loop_invariant,
                        loop_invariant_redundantly,measured_by,measured_by_redundantly,maintaining,
                        maintaining_redundantly,maps,maps_redundantly,\\new_objects,old,refining,represents,
                        requires,set,signals,signals_only,\\such_that,unreachable,when,working_space,expand,witness,instantiate,auto},%
        morekeywords = [6]{%
                        abrupt_behavior,abrupt_behaviour,also,axiom,behavior,behaviour,
                        constraint,exceptional_behavior,exceptional_behaviour,initially,
                        invariant,model_behavior,monitors_for,normal_behavior,normal_behaviour,readable,writable},%
        morekeywords = [7]{%
                        \\seq_empty,\\seq_def,\\seq_singleton,\\seq_get,\\seq_put,\\seq_reverse,\\seq_length,\\index_of,\\seq_concat,\\empty,\\singleton,\\set_union,\\intersect,\\set_minus,\\all_fields,\\infinite_union,\\strictly_than_nothing},
        morekeywords=[8]{@param,@returns,@preconditions,@postconditions,@invariants},%
        morecomment = [s]{/*}{*/},%
        morecomment = [l]//,%
        morecomment = [n]{(*}{*)},%
        moredelim =*[s]{/*@}{@*/},%
        moredelim =*[s]{/*+KeY@}{@+KeY*/},
        moredelim =*[l]//@,%
        moredelim =*[l]//+KeY@,
        morestring=[b]",%
        morestring=[b]',%
        sensitive,%
}[keywords,comments,strings]

\lstdefinestyle{color}{
  basicstyle=\upshape\ttfamily\small,%
	commentstyle=\slshape\small\color{black!40},
        keywordstyle=\bfseries\small\color{purple},
        keywordstyle=[4]\small\color{blue},
	keywordstyle=[5]\small\color{red},
	keywordstyle=[6]\small\color{red},
        keywordstyle=[7]\small\color{blue},
	numberstyle=\tiny,
}

\lst@Key{numbers}{none}{%
    \def\lst@PlaceNumber{\lst@linebgrd}%
    \lstKV@SwitchCases{#1}%
    {none:\\%
     left:\def\lst@PlaceNumber{\llap{\normalfont
                \lst@numberstyle{\thelstnumber}\kern\lst@numbersep}\lst@linebgrd}\\%
     right:\def\lst@PlaceNumber{\rlap{\normalfont
                \kern\linewidth \kern\lst@numbersep
                \lst@numberstyle{\thelstnumber}}\lst@linebgrd}%
    }{\PackageError{Listings}{Numbers #1 unknown}\@ehc}}

\lst@Key{linebackgroundcolor}{}{%
    \def\lst@linebgrdcolor{#1}%
}
\lst@Key{linebackgroundsep}{0pt}{%
    \def\lst@linebgrdsep{#1}%
}
\lst@Key{linebackgroundwidth}{\linewidth}{%
    \def\lst@linebgrdwidth{#1}%
}
\lst@Key{linebackgroundheight}{\ht\strutbox}{%
    \def\lst@linebgrdheight{#1}%
}
\lst@Key{linebackgrounddepth}{\dp\strutbox}{%
    \def\lst@linebgrddepth{#1}%
}
\lst@Key{linebackgroundcmd}{\color@block}{%
    \def\lst@linebgrdcmd{#1}%
}

\newcommand{\lst@linebgrd}{%
    \ifx\lst@linebgrdcolor\empty\else
    \rlap{%
        \lst@basicstyle
        \color{-.}%
        \lst@linebgrdcolor{%
        \kern-\dimexpr\lst@linebgrdsep\relax%
        \lst@linebgrdcmd{\lst@linebgrdwidth}{\lst@linebgrdheight}{\lst@linebgrddepth}%
        }%
    }%
    \fi
}
\makeatother

\usepackage[shortlabels,inline]{enumitem}
\usepackage{subcaption}     %
\usepackage{siunitx}        %

\newcommand{\enquote}[1]{``#1''}

\newif\ifShowDiff

\newcommand{\diff}[1]{\ifShowDiff\textcolor{blue}{#1}\else#1\fi}
\newcommand{\diffComment}[1]{\ifShowDiff\textcolor{blue}{#1}\fi}

\begin{document}

\title{A New Interaction Concept for Interactive and Autoactive Program Verification}

\author{Wolfram Pfeifer}
\email{wolfram.pfeifer@kit.edu}
\orcid{0000-0002-9478-9641}
\affiliation{%
  \institution{Karlsruhe Institute of Technology}
  \city{Karlsruhe}
  \country{Germany}
}

\author{Mattias Ulbrich}
\email{ulbrich@kit.edu}
\orcid{0000-0002-2350-1831}
\affiliation{%
  \institution{Karlsruhe Institute of Technology}
  \city{Karlsruhe}
  \country{Germany}
}

\author{Daniel Drodt}
\email{daniel.drodt@tu-darmstadt.de}
\orcid{0000-0003-3036-8220}
\affiliation{%
  \institution{Technical University of Darmstadt}
  \city{Darmstadt}
  \country{Germany}
}

\begin{abstract}
  Fully functional program verification is an undecidable---and,
  hence, inherently difficult---task, that is not automatically
  solvable but typically requires user interaction and guidance.
  Existing verifiers either work autoactively, requiring the user to
  write annotations in source code, without the possibility to inspect
  the proof state or intervene in case of an unsuccessful attempt, or
  allow interactions on a logical encoding that is on a lower level
  than the user-provided specifications.

  We present a novel interaction concept which allows the user to
  inspect and interact with the proof state on source code and
  specification level.  This minimizes the mental gap between the
  representations.  We provide an implementation of the concept as a
  plugin for the Java verification engine \KeY, and show with a user
  study that this prototype can be beneficial for users to understand
  the proof state and find defects in source code or specifications.
\end{abstract}


\maketitle

\section{Introduction}
\label{sec:intro}

Deductive software verification has made substantial pro\-gress in
recent
years~\cite{PKGLSWESBMP2025,OHJv2020,SOH2021,BSUWW2024}, yet its adoption in
industry remains limited~\cite{tCCG2024}. Verifying non-trivial
programs written in real-world programming languages is inherently
difficult, and---because the general problem is undecidable---this
difficulty will persist even as tools improve.

Deductive verification is commonly organized modularly: functions and
loops are annotated with specifications (contracts, loop invariants,
etc.), and proofs for program correctness are constructed from these
component specifications. Successful verification is therefore
typically an iterative process: it rarely succeeds on the first
attempt. Instead, a verification engineer must inspect remaining proof
obligations, determine why a proof attempt failed, and take corrective
action. Broadly speaking, there are three reasons a proof may not
succeed:
\begin{enumerate}
\item the implementation is incorrect,
\item the specification is incorrect or too weak,
\item the tool’s automation is insufficient to close the proof.
\end{enumerate}

The appropriate response differs between these cases. In the first
two, the source code or the specification (or both) must be adapted to
resolve the mismatch. In the third case, deeper insight into the open
proof goal is required to diagnose why the automation cannot close it
and to decide how to intervene. In all cases, understanding of the
proof goal requires expert knowledge of both the program under
verification and the verification tool.

In this paper we present an integrated interaction concept that
supports users in assessing proof situations and interacting with
proofs at the source code level. The central idea is to present all
available verification information as overlays on the program source
and to enable interaction directly at that level.
Consequently, proofs that previously required
considerable low-level manual intervention can often be discharged by
iteratively guiding automation without leaving the specification
level. When necessary, users may still perform fine-grained manual
interaction on more detailed abstraction levels.
Conceptually, \diff{proof} guidance given interactively should best be
recorded as proof scripts on the same abstraction level to allow
replay and persistence of manual prover guidance.

We implemented this interaction concept by combining interactive and
autoactive verification on top of the deductive Java verification
engine \KeY~\cite{ABBHSU2016} which is founded on a program logic in
which inference rules are applied at a fine granularity. The base
logic is \diff{an extension of} first-order logic tailored to model
Java states; proofs are browsable and interactable in the \KeY user
interface. This detailed representation exposes many details of the
symbolic state of the open goal and hence allows for subtle proof
interactions, but it can also overwhelm users quite easily: It is
often hard for a \KeY novice to understand why a particular formula
appears in the proof obligation and how it relates to the source
program.

The proposed interaction concept addresses this gap between program
and proof by (i)~lifting information from the technical proof
presentations back to the source code level, (ii)~providing visual
references between source code level statements and their counterparts
in the proof, and (iii)~enabling source code level invocation of proof
tactics and of inference rules on the detailed level and the recording
of proof scripts. The overall goal is to reduce the cognitive effort
required to switch between the source code level mental model and the
detailed logical proof model, thereby making deductive verification
more accessible and productive.

The presented integration is particularly well suited for introducing
programmers to formal reasoning.  Specification languages are
typically designed to be familiar to practitioners of the specified
language. Much of the Java Modelling Language (JML)~\cite{LBR1999},
e.g., is designed to be comprehensible to a Java developer with basic
first-order-logic training. By contrast, understanding failed proof
attempts at the logical level often requires more specialized training
in logic and proof calculi. The approach presented here enables users
to transition gradually from high-level, specification-level reasoning
to the detail-rich logic backend when required. By narrowing the gap
between \diff{prover} and specification, interacting with the prover
at the specification level offers a promising way to reduce the
learning curve for practitioners.

The implemented prototype of this proof presentation and interaction
concept was evaluated in a small user study with \KeY experts. The
results indicate that presenting proof information inline in the
source reduces cognitive overhead and can speed up some verification
tasks, even for users more experienced with the traditional sequent
view.

The paper is structured as follows: \autoref{sec:background} contains
the necessary background about JML and \KeY.  In
\autoref{sec:ui-source-level}, we present the idea of user interaction
on source code level, the central hypothesis, and our concept of what
user interaction should look like, independent of the concrete prover
it is implemented in.  \autoref{sec:impl} refines these ideas and
explains how they are implemented prototypically in the \KeY system,
as well as the challenges that had to be solved.  The user study
described in \autoref{sec:user-study} demonstrates that our ideas can
be beneficial for interactive verification.  Related work is shown in
\autoref{sec:related-work}, before we conclude with a short summary
and future work ideas in \autoref{sec:conclusion}.

\section{Background} %
\label{sec:background}
While elements of the interaction concept are not specific to Java and
can be applied to other imperative languages, we present it as an
extension of existing technologies using the \emph{Java Modeling
  Language} as specification language and the Java verification tool
\emph{\KeY} as the verification engine.

\subsection{The Java Modeling Language}
\label{sec:jml}

The Java Modeling Language (JML)~\cite{LCN2022}, %
is a behavioral interface specification language for Java and the de facto standard for formal specifications in Java. Following the \enquote{design by contract} paradigm~\cite{M1992}, it supports
modular program specification where each method is annotated with a
\emph{method contract}.
\autoref{lst:swap} exemplarily demonstrates the parts of a JML
contract. JML is always written in a comment with \lstinline|@| as the
first character. Here, we declare a \enquote{normal behavior} case,
i.e., no exception is thrown, with the precondition
(\lstinline|requires|) that the parameters \lstinline|i| and
\lstinline|j| are valid indices of the array~\lstinline|a| and the
postcondition (\lstinline|ensures|) that \lstinline|a|'s values at
indices \lstinline|i| and \lstinline|j| are now swapped while all
other values are unchanged. Note that JML expressions are Java
expressions extended with universal (\lstinline|\forall|) and
existential (\lstinline|\exists|) quantification, implication
(\lstinline|==>|), and an \lstinline|\old($e$)| construct which
evaluates the JML expression $e$ in the state at the beginning of the
current method call, among others. We call the parts of a JML
specification, e.g., the precondition, \emph{clauses.}
The \lstinline|assignable| clause specifies the frame condition, i.e.,
the set of \emph{heap locations}
that can be modified by the method.

\begin{lstlisting}[language={[JML]Java}, caption={JML specification of a \lstinline|swap| method. More concise specification is possible.}, label=lst:swap, float=tb]
/*@ public normal_behavior
  @ requires 0 <= i < a.length && 0 <= j < a.length;
  @ ensures a[i] == \old(a[j]) && a[j] == \old(a[i]);
  @ ensures (\forall int k; 0 <= k < a.length;
  @   k != i && k != j ==> a[k] == \old(a[k]));
  @ assignable a[*];
  @*/
public void swap(int[] a, int i, int j) {/*...*/}
\end{lstlisting}

Such a formal contract as the one for \lstinline|swap| serves not only to
document the method's expected behavior precisely, it can also be used
as input tools like OpenJML~\cite{C2011}, VerCors~\cite{BH2014}, and \KeY
to verify (or disprove) the implementation's correctness with regard
to the contract.

JML also permits annotations in a method body. Particularly relevant
for this paper are assertions and assumptions both used to indicate
that an expression is supposed to hold whenever the respective line
with the statement is reached. But there is \diff{a} difference, as
\autoref{lst:assert} shows: Whereas the assumption \lstinline|n >= 0|
in the first line of \texttt{m} is silently assumed to hold without
proof, the assertion for \lstinline|n > 0| in the third line raises a
corresponding proof obligation.  Assertions and assumptions may
contain JML expressions, e.g., quantified formulas.
The purpose of such annotations are, again, both for documentation but
more importantly for prover guidance since they introduce local lemmas
that need to be proved locally, but that also can make the proof of
surrounding proof obligations easier.

\begin{lstlisting}[language={[JML]Java}, caption={JML assumptions and assertions.}, label=lst:assert, float=tb]
int m(int n) {
  //@ assume n >= 0;
  n++;
  //@ assert n > 0;
  return n;
}
\end{lstlisting}

\subsection{The \KeY Tool}
\label{sec:key}

\KeY~\cite{ABBHSU2016,BBDHLPUW2025} is a deductive verification tool
for Java programs based on the dynamic logic JavaDL, a program-logic
variant of modal first-order logic with Java programs embedded in
modal operators. It introduces state-dependent \emph{program
  variables} and two modal operators: the \enquote{diamond}
$\dia{p}{\phi}$ and the \enquote{box} $\dlbox{p}{\phi}$, where $p$ is
a fragment of a Java program and $\phi$ is a JavaDL
formula. Intuitively, the formula $\dia{p}{\phi}$ holds in a state iff
$p$ terminates (without exceptions) when started in state~$s$ and, in
the poststate after $p$, $\phi$ holds; $\dlbox{p}{\phi}$ is the
partial operator, it holds when $\dia{p}\phi$ holds or when $p$ does
not terminate.

A proof obligation for the \diff{total} correctness of a
method~\java{m} \diff{w.r.t.\ precondition~$\pre$ and postcondition
  $\post$} can be expressed in JavaDL in the form
$\mathit{pre}\rightarrow\dlf{res=m();}{\mathit{post}}$.

To prove a JavaDL formula $\phi$ correct, \KeY relies on a sequent
calculus and shows that the sequent $\seq{}{\phi}$ can be derived. Its
calculus includes the familiar propositional and first-order rules,
rules for reasoning about boolean values and integer arithmetics, as
well as rules to symbolically execute the program fragment in
modalities. The rule $\ruleName{ifElseSplit}$ below matches when an
if-else-statement is encountered and splits the proof goal into two
branches: one, where the if's condition holds and the then-branch is
executed; one where the condition does not hold and the else-branch is
executed. Thereby, all cases are covered. For details of these rules
and their implementation, we refer to~\cite{BBDHLPUW2025}.
\[
  \seqRule{ifElseSplit}
  {\sequent{b\keyeq\keybooltrue}{\dia{s_1}{\phi}}\\
  \sequent{b\keyeq\keyboolfalse}{\dia{s_2}{\phi}}}
  {\sequent{}{\dlf{if ($b$) $s_1$ else $s_2$}{\phi}}}
\]

These symbolic execution rules simplify and evaluate the code
in modalities piecemeal until all modalities have been fully executed
and only first-order formulas remain.

In addition to sequent rules like $\ruleName{ifElseSplit}$, \KeY
offers rewrite rules such as $t\keyeq t\rightsquigarrow\keytrue$,
which can be applied to terms and (sub-)formulas in the sequent. Rules
can also have additional conditions, such as a formula $\phi$ being
present in the antecedent. This greatly increases the calculus'
flexibility and expressiveness. For details of the realization of
these rules in the form of \enquote{taclets,} see~\cite{RU}.

To model state changes in a total, structured manner, JavaDL also
introduced \emph{updates.} An elementary update $\java{x}\upd t$
represents the state change where variable~\java{x} now has the value
of term~$t$. Updates can be composed to parallel updates, e.g.,
$\java{x}\upd\java{y}\parupd\java{y}\upd\java{x}$, which represents
swapping the values of variables $\java{x}$ and $\java{y}$, and can be
applied to terms and formulas, e.g.,
$\upl\java{n}\upd\java{n}+1\upr(\java{n}>0)$.

By relying on JavaDL formulas, \KeY avoids the typical approach of
translating the program under verification to a lower-level
intermediate language or SMT problem, which is then verified. Instead,
the program is directly embedded in modalities.

\begin{figure}
  \centering
  \includegraphics[width=\linewidth]{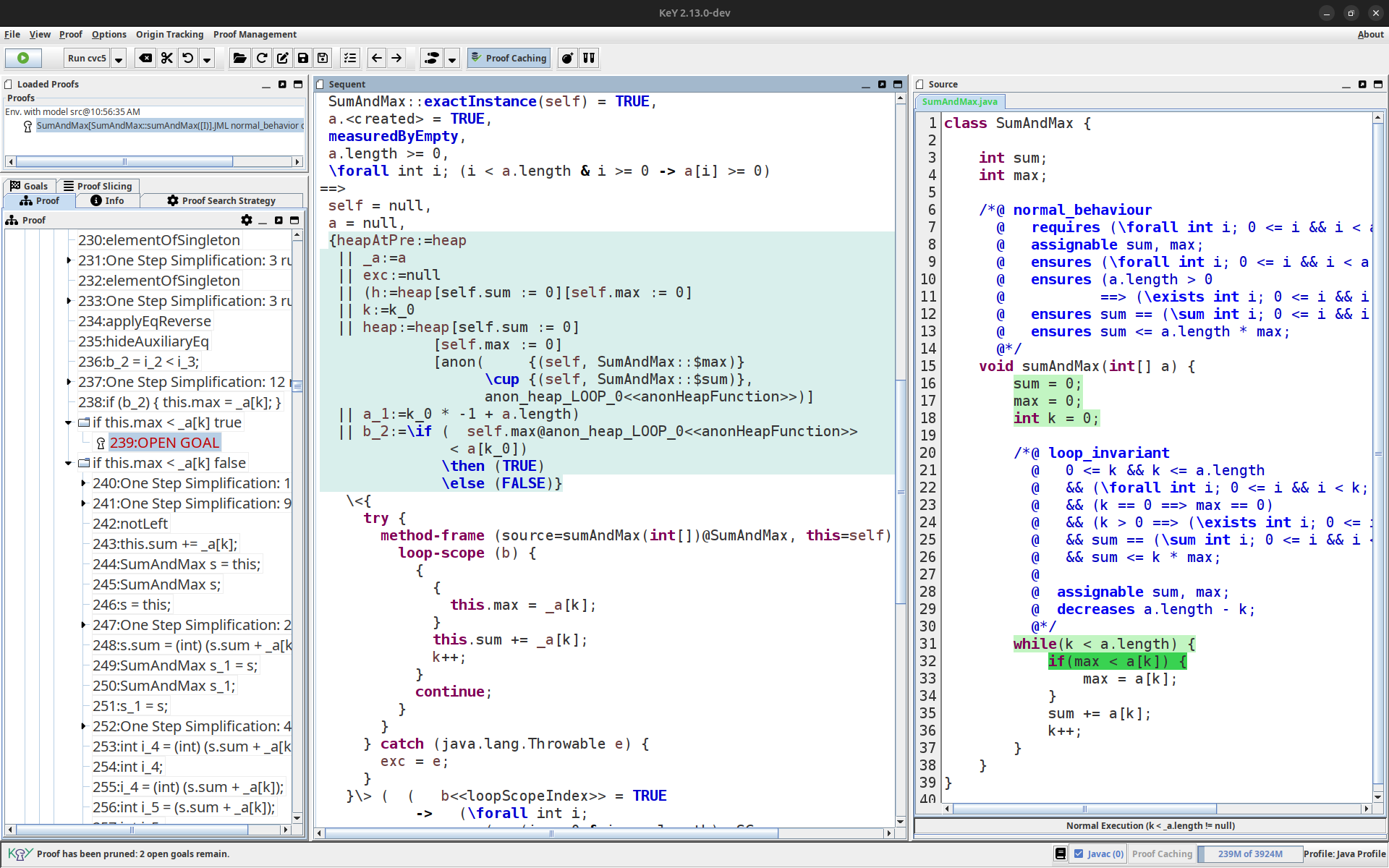}
  \Description{A picture of the \KeY GUI.}
  \caption{The \KeY GUI. On the left is the proof tree, in the center
    the sequent view, and on the right, the code under verification.}
  \label{fig:ui}
\end{figure}

\KeY provides a GUI with the sequent shown as an explicit \emph{goal
  state}---an, as yet, unproven sequent in the proof state---where
a user can manually apply any applicable rules directly on selected
terms and formulas, permitting a great deal of control and
fine-grained interaction, see \autoref{fig:ui}.
At the same time, \KeY has a powerful automatic proof mode that is
able to discharge many proof obligations stemming from non-trivial
properties of non-trivial programs.
This combination of interaction patterns enabled verification of
complex, real-world Java programs such as TimSort~\cite{DRDBH2015},
\lstinline|LinkedList|~\cite{HMBdvd2020},
\texttt{IdentityHashMap}~\cite{ddKJUW2023} and
ips\textsuperscript{4}o~\cite{BSUWW2024}. Nevertheless, manual
interactions require a great deal of expertise with the program under
verification as well as JavaDL and \KeY's calculus.

As a middle ground between full automation and manual interaction,
\KeY provides an array of \emph{macros,} enabling more focused
automatic rule applications. With macros, one can, e.g., execute only
symbolic execution rules or only propositional simplification. This is
especially helpful when attempting to understand a given goal state to
uncover why verification fails.

\section{User Interaction on Source Code Level}
\label{sec:ui-source-level}

In this section, we describe our concept for interaction.  While we
use \KeY in the motivation to demonstrate shortcomings of interaction
on a low-level logical representation, the concept and ideas in
this section are independent.

\KeY's current interaction pattern typically proceeds as follows: A
method is chosen for verification in the GUI, a JavaDL proof
obligation is generated from the method's JML specification and
presented as an initial sequent. Then, the proof automation is started
to attempt verification, with possible interactive steps in between to
close some goals beyond the prover's capabilities.

This approach of expressing a proof obligation in JavaDL formulas
and proving these formulas with a fine-grained, precise
calculus provides some immediate benefits. No complexity is hidden and
a user can fully inspect the proof steps and how they changed or
introduced proof goals. Further, since the formulas may contain Java
code directly, there is a narrower gap between the method to be
verified and the initial proof obligation than achievable with
translation to an intermediate language or SMT.

However, these benefits come at the cost of some complexity: Even the
initial proof obligation contains elements not present in the original
program. For instance, since JavaDL is more expressive than Java, we
add preconditions to restrict possible values of method parameters and
states of the heap. This discrepancy is compounded by the non-trivial
changes to the sequent introduced by rule applications: To model
Java's complex semantics accurately, programs are significantly
altered during symbolic execution. Formulas are then further
simplified, moved around, discarded as irrelevant, etc. While each
step is inspectable and comprehensible, understanding the relation of
a sequent to the program after thousands or tens of thousands of rule
applications is non-trivial. \diff{Similar problems arise from
  encoding a program in intermediate languages or SMT, as implicit
  knowledge of the target language is added to the translation,
  increasing the gap to the source code.}

From the case studies of more complex algorithms and data structures
completed in recent years~\cite{BSUWW2024,S2023,ddKJUW2023}, we
developed the following hypothesis: \emph{To minimize the mental gap,
  all interactions with the prover should be as
  close as possible to the original user input, i.e., the specified
  source code.}

From this central hypothesis, we derive the following points for our
concept.  First, it is important that users can orient themselves in
the proof and understand what has to be proven in the current proof
branch (respectively goal).  Second, open goals should be presented
inline in the source view, in a language and form as close as possible
to the original user input.  Third, the user should be able to perform
interactions to inspect and manipulate directly on this source view;
ideally they should be able to finish the proof without the need to
descend to the underlying logical level of the prover.  This also
means that not all intermediate steps of the underlying prover need to
be inspectable and manipulatable by the user to be able to understand
the required actions in the proof.  Instead, it is sufficient to see
the goals in specific key points in the proof, while relying on the
automatic facilities of the system to fill in the intermediate steps.

Although we apply this approach to \KeY and the hypothesis is built on
experience in the \KeY system, we maintain that our conclusion
generalizes: Any verification tool where proving occurs on a lower
level then the source code level must strive to minimize the gap
between the two. Presenting the remaining goals in a source code level
view is a general approach, regardless of the concrete verification
tool and methodology.

\subsection{The Novel User Interaction Concept}
\label{sec:concept}

Our interaction concept implements a novel approach to present a
specific open goal state to the user.  The formulas from the internal
logical proof engine should be presented as assertions and assumptions
in the source code. Assumptions are to represent knowledge about the
program collected in the current goal. Assertions ought to signify
formulas that remain to be proven. We must take care to present the
formulas as close as possible to their original formulation as given
by the user.  This minimizes the mental gap and allows the user to
investigate the current proof situation without needing to know about
the underlying logical encoding: It is sufficient to understand the
program and its specification. \autoref{lst:jml-goal-state}
demonstrates the desired result of this approach.
\diff{The use of assumptions and assertions is motivated by the fact
  that these are already present in JML and represent a known and
  intuitive concept for the user.  However, since they are only used
  for ``capturing'' the current state of the prover and presenting the
  goal, and in particular do not represent assumptions on the input
  level, they need to be clearly distinguishable from user-given
  input.  In \autoref{lst:jml-goal-state}, this is accomplished with a
  gray background.}

\begin{lstlisting}[float, caption={The desired output of the new user interaction pattern. The assertions and assumptions with the gray background represent the current proof goal. The method \java{zero} sets all elements of integer array \java{a} to zero. Verification fails, as the precondition \java{to < a.length} is missing; hence, the \diff{second} assertion cannot be shown \diff{(\lstinline|a[k]| may be out of bounds)}. \diffComment{LST CHANGED!!!!!!}}, label={lst:jml-goal-state}, linebackgroundcolor={
        \ifnum\value{lstnumber}>13
            \ifnum\value{lstnumber}<19
                \color{gray!10}
            \else
                \ifnum\value{lstnumber}>20
                    \ifnum\value{lstnumber}<23
                        \color{gray!10}
                    \fi
                \fi
            \fi
        \fi}]
/*@ public normal_behavior
  @ requires 0 <= to;
  @ assignable a[0..to];
  @ ensures (\forall int i; 0 <= i < to; a[i] == 0);
  @*/
void zero(int[] a, int to) {
  int j = 0;
  /*@ loop_invariant 0 <= j <= to
    @ && (\forall int k; 0 <= k < j; a[k] == 0);
    @ assignable a[0..to];
    @ decreases a.length - i;
    @*/
  while (j < to) {
    //@ assume 0 <= to;      // precondition
    //@ assume j < to;       // loop guard
    //@ assume 0 <= j <= to; // loop invariant ...
    //@ assume (\forall int k; 0 <= k < j; a[k]==0);
    //@ assert 0 <= j < a.length; // a[j] in bounds
    a[j] = 0;
    ++j;
    //@ assert 0 <= j <= to; // loop invariant ...
    //@ assert (\forall int k; 0 <= k < j; a[k]==0);
  }
}
\end{lstlisting}

\diff{Note that with our concept, we only aim to show a single proof
  goal (that is, a single leaf of the proof tree) at the same time.
  The rationale behind this is that while it might be possible to
  display multiple goals at once, it would become unreadable already
  for very small programs.
  However, it is possible to inspect different goals: When the user
  navigates in the proof tree to a suitable goal, i.e., where symbolic
  execution and update simplification is completed, the view is
  updated accordingly.}

By presenting the goal state inline together with (relevant parts of)
the initial program, we can provide a basis for investigating the goal
state. We should further aid the user by highlighting the source of the
annotations. For instance, hovering over the assumption %
\lstinline|0 <= to| ought to highlight the \lstinline|requires|
clause, as this is the formula's origin.

To further minimize the gap between proof and specification, it is
imperative that the terms and formulas are as close as possible to the
original input. A verifier might, for example, normalize integer
functions and relations to improve automation: \lstinline|0 <= to|
would be transformed to the term $\java{to} > -1$. While this is
necessary for the prover, it hinders the user, especially as these
normalizations compound. Hence, we must ensure that terms and formulas
are represented closely to their \emph{original} form.

In the same vein, we want to purge \emph{implicit} information from
the new representation. As stated above, provers often add implicit
preconditions, e.g., restricting the state to possible Java heaps, the
value range of integer parameters to the range of Java's
\lstinline|int| type, or ensuring that relevant objects have been
created. These formulas have no equivalent on the specification level,
as they are already implicit on that level, and they are only
necessary for proving but rarely for goal
comprehension. Thus, they must not be added as
assumptions.%

In program verification, formulas are evaluated in a context, that is
the current assignment of local variables and heap state.  In our
concept, this is represented by the positioning of assertions and
assumptions: They are inserted into the source code in the position
corresponding to the state where they have to be proven respectively
are known to hold.

However, this is insufficient, since a single formula might relate
different program states.  Fortunately, JML provides a tool for this:
the \lstinline{\old} operator.  In addition to the standard form,
which only has a term as single parameter and evaluates this term in
the pre-state of the method, there is also a second form available.
This second form has a (Java) label as a second parameter, and
evaluates the given term in the state at the label.  With this
operator, it is possible to insert a formula in a specific line, and
still refer to states at different positions in the code.
\diff{\autoref{lst:jml-example-old} shows an example where the
  standard {\lstinline{\\old}} without label is not sufficient and the
  variant with label is needed.  However, note that this requires to
  introduce the additional label into the source code.}

\begin{lstlisting}[float, caption={Example of the use of the operator
{\lstinline{\\old}} with label. The first assertion refers to two
states that are both not the pre-state. In contrast to the second
assertion, the standard operator without label is not
sufficient. \diffComment{LST added. Please check! Is this too
  artificial?}}, label={lst:jml-example-old}, linebackgroundcolor={
\ifnum\value{lstnumber}>4 \ifnum\value{lstnumber}<7 \color{gray!10}
  \fi \fi}]
//@ ensures a[i] == \old(a[i]) + 8;
void inc(int[] a, int i) {
  a[i] = a[i] + 1;
  l: add(a, i, 7);
  //@ assert a[i] == \old(a[i], l) + 7;
  //@ assert a[i] == \old(a[i]) + 8;
}
\end{lstlisting}

We silently assume that there is at most one statement per
line. Otherwise, the placement of assumptions would be difficult to
compute and unnecessarily difficult to read. Also, return statements
must be of the form \lstinline|return x;|, where \java{x} is a
variable or constant, since we cannot insert assertions after the
return.

To realize this concept, \diff{the verifier must offer some
  additional functionality}. First, we must be able to track the
\emph{origin} of terms and formulas, i.e., whether they come from the
precondition, a loop invariant, etc. Second, we must be able to
adequately translate terms and formulas from the logical level back to
the source code level.

\subsubsection{Tracking Origins of Terms and Formulas}
\diff{To track the origin of terms and formulas, they need to be
  associated with an \emph{origin reference}. Such a reference must
  entail a type (e.g., $\mathit{requires}$), file information, and
  line and column ranges.}

Implicitly added formulas \diff{must be} assigned a special
\enquote{implicit} type to differentiate them from explicit
formulas. Further, they \diff{can} have no file and line/column
information, since they do not stem from the source code. %

One might expect that origins are attached to formulas, as they are
typically to what \diff{specification} clauses are translated.
However, this tends to break when formulas are split, for instance a
conjunction into their conjuncts, which automatic provers must often
do.
Therefore, we assign origins to the lowest entity possible, the
operator, i.e., the $\wedge$ in $\phi\wedge\psi$.
If no origin is found for a formula, the subterms are searched recursively
for origins and their union is shown.

Once the origins have been assigned to the operators of the initial
proof obligation, the verification tool must ensure that they are
maintained, so as to still be available for open goals after the proof
attempt failed. We discuss how we solve this for \KeY in
\autoref{sec:impl}.

\subsubsection{Translating Terms Back to Source Code Level}

With the origin information in place, the next step is to translate
the formulas from the underlying logical representation (e.g., a
sequent) back to JML.  For this, the formulas are classified according
to their origin type as either \textit{implicit}, \textit{assume} or
\textit{assert}.  Implicit formulas are not shown in the source code
view.  Formulas of the other two categories are shown correspondingly
as JML \lstinline|assert| or \lstinline|assume| statements.

For the formulas in JML statements, it is important to stay as close
as possible to the user input.  Therefore, if the formula on the
logical layer has a single unique origin, the original string from the
user input should be displayed verbatim.  This \enquote{reverts} all
normalization steps that have been done on the logical level, for
instance rewriting \diff{to} negation normal form, polynomial normal
form, etc.  However, often a formula has been transformed
significantly, for example by applying equations or combining multiple
input formulas, and a direct mapping to a JML formula is not possible.
In this case, a back-translation has to be applied, which maps
operators of the logical level back to their corresponding operators
in JML.

In addition to the translation of formulas, also the position where to
show them in the source code has to be determined.  This is, however,
highly dependent on the concrete representation of state in the logic
and thus on the prover the concept is implemented in.  For \KeY, we
discuss a solution in \autoref{sec:impl}.

Even when all the technical requirements are met, there remains a
crucial conceptual limitation: The tracking of origins and translation
back to the source code level is only possible if there is only one
symbolic execution.
Although provers might be capable of verifying relational
properties---for instance, \KeY can reason about information flow and
dependency contracts~\cite{BKU2015,ABBHSU2016}---their proof
obligations contain multiple programs to execute, making translation
into a single Java program impossible.  Nevertheless, verification of
functional correctness remains the most popular use case of \KeY and
similar tools, reinforcing the relevance of our approach.

\subsection{Advancing Proofs and Proof Scripts}

Representing information in open proof goals in terms of the original
source code and cross-referencing them to their logical counterparts
are vital prerequisites for enabling the comprehension of proof
obligations in program verification. However, presenting proof goals
well is \diff{insufficient} to address how proofs are conducted at
that level; allowing and supporting user interaction to advance the
proof is an equally important aspect of the process.

The interaction style that has, traditionally, proved particularly
successful in \KeY is a point-and-click paradigm, in which the user
selects the rule to be applied from a context-sensitive list of
applicable inference rules. This approach provides fine control over
the proof process while sparing the user from having to memorize the
names of the c. 1\,500 rules implemented in \KeY.

In comparison, in classical autoactive verification, interaction is
achieved by composing verification-only statements and similar support
annotations in the program source code guiding the prover and helping it
close obligations that cannot be discharged fully automatically.

In the novel interaction concept, we intend to combine the benefits of
both approaches: To allow high-level user interaction in terms of
the surrounding source code, but to allow at the same time to revert
to the technical level if needed.

\diff{The new interaction pattern aims to lift a number of
  interactions that are already supported on the logical level to the
  source code level. After finishing symbolic execution of the proof
  obligation (via a number of different macros), running the automatic
  prover is one of the coarse grained interactions lifted from the
  technical to the source code level. Further possible} proof
interactions to guide the prover towards closing a proof include
splitting of conjunctively connected proof obligations into parts,
introducing case distinctions (so-called cuts), and instantiating
quantifiers with JML expressions. \diff{For improved clarity, it must
  also be} possible to drop individual displayed assumption formulas
from the goal state if the user deems that they do not contribute to
the proof but might rather cause the automation to go astray.

\diff{We retain the possibility of fine-grained user interaction
  mechanisms, such as applying elementary proof steps (inference
  rules) directly on the logic level,} without any abstractions to the
source code level. Although this fall-back is intended as a last
resort for situations where higher-level interaction is insufficient,
experience shows that there are genuine cases in which the solution
requires interaction at \diff{the logical level, e.g., JavaDL}. In
such cases, technical details of the surrounding verification
condition that are not visible at the source level can be crucial for
the reasoning. A typical example arises when one must reason about the
\enquote{createdness} status of objects. Createdness is not exposed at
the JML level---by design, as JML assumes that all referable objects
are created---whereas the logical level can express finer
distinctions.

\begin{lstlisting}[label=lst:proof-script,float,caption={Vision how a typical interaction in the \KeY GUI---expanding an invariant, obtaining a Skolem constant, instantiating a quantifier with it, and closing the proof with the auto mode---could be persisted as a JML script.}, showstringspaces=false, basewidth=.48em]
/*@ assert (\exists int i; 0<=i<a.length; a[i]>0) \by {
  expand on="\invariant_for(this)";
  witness "(\exists int j; 0<=j<a.length; a[j]>1)"
      as="j_0";
  instantiate "(\exists int i; 0<=i<a.length; a[i]>0)"
      inst="j_0";
  auto;
} @*/
\end{lstlisting}
The extensibility of proofs via mouse interactions\diff{, whether at the
more abstract source code level or at the more technical level,} is
very user-friendly and allows very targeted proof progress in many
cases without lengthy text inputs. A drawback of this form of
interaction is that it is little persistable and is sensitive to small
changes in the program or the specification. Because it is present as
user input in a concrete proof, it can hardly be adapted to changed
situations. The application to even slightly different proof trees is
challenging~\cite{BK2004}. The solution foreseen in the concept is the
transcription of user interactions as proof scripts at the code
level. Autoactive provers support annotating assertions in code with
commands for guiding the prover to address similar
challenges. Dafny~\cite{L2010}, for example, has annotations that can
be attached to assertions to provide guidance on how proofs should be
conducted. \KeY likewise has a \diff{new and experimental}
mechanism,\footnote{See
  \url{https://keyproject.github.io/key-docs/user/ProofScripts/jml/}
  for details on the JML script mechanism in \KeY} with which proofs
can be annotated at JML assertions. The abstraction level in JML
scripts corresponds to that of interactions at the source code level.

It is therefore natural that user interactions (on both levels) should
be recorded as commands in JML proof scripts to make them accessible
for users, and to make them available in a persistent manner for
automatic proof replay. \diff{Thus our concept brings point-and-click
  interaction to a source code level view and extends the expressivity
  of the source code, permitting rule applications, cuts, macros,
  etc. as JML script commands.  For instance, a typical interaction in
  \KeY is that first an invariant is unrolled, a universal quantifier
  from this invariant is skolemized resulting in a new constant, and
  then an expression containing this constant is provided as an
  instantiation for an existentially quantified formula. Afterwards,
  the proof is found automatically by the proof search strategy.
  \autoref{lst:proof-script} show how such an interaction in the GUI
  could be represented (potentially automatically) as a
  script.}%

\diff{This naturally allows for a new \KeY workflow: The user writes
  JML commands which advance the proof, introduce new goals and close
  some others. If the proof does not close completely, the new view
  allows low-effort inspection of the proof state, which leads to
  refinement of the proof script. The cycle continues until the proof
  succeeds. Examination and interaction of the logical level remains
  optional in as many cases as possible.}

\section{Prototype Implementation of the Concept in \KeY}
\label{sec:impl}
\begin{figure}[t]
  \includegraphics[width=\linewidth]{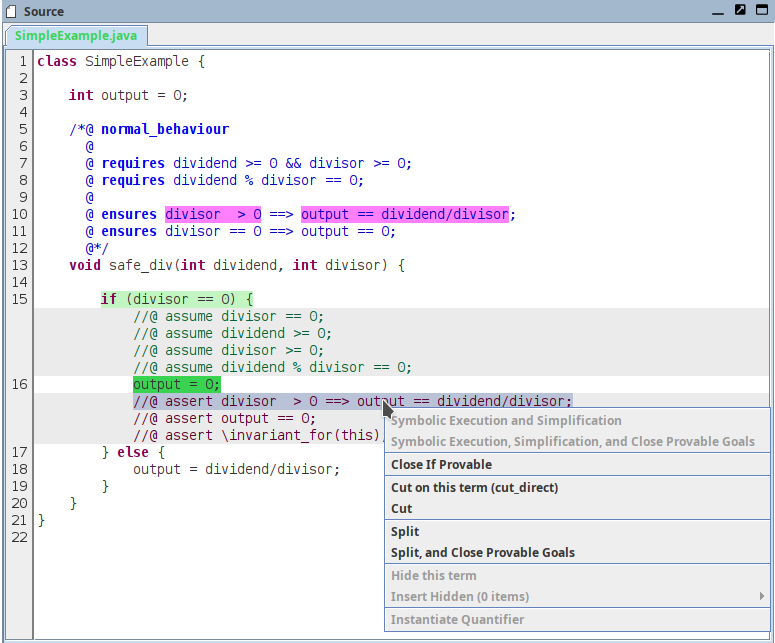}
  \Description{A Java file annotated with JML comments. The currently
    active statement is highlighted in green, above it are JML assume
    statements, below JML assert statement. The active statement and
    its annotations are surrounded by an if-else-statement in the
    method body of safe\_div.}
  \caption{Screenshot of the prototype
    implementation. \diffComment{Changed: Added context menu
      screenshot}}
  \label{fig:screenshot}
\end{figure}
The interaction concept outlined in \autoref{sec:ui-source-level} has
been implemented in a prototype plugin extension\footnote{The
  implementation can be found at
  \url{https://github.com/KeYProject/key/tree/pfeifer/sourceInteractionPrototype}}
for the graphical user interface of \KeY comprising some 7\,000 lines
of Java.
A screenshot of the GUI can be seen in \autoref{fig:screenshot}.
While still early in development, the prototype contains the following
features.  First, the symbolically executed lines for the current goal
are shown with green background, with the most recently executed line
more saturated.  A prerequisite for this is that the symbolic
execution has been completed, and no state transitions in form of
modalities or updates are present on the sequent.  Since typically the
symbolic execution is done fully automatic, and steps in between are
not interesting for the user, this is not a hard restriction.  Second,
as described in \autoref{sec:concept}, the generated JML assumptions
and assertions that represent the current proof goal are shown inline
in the source view.  These insertions for the current goal can be
distinguished easily from the user-provided specification by the gray
background and the missing line numbers.  Note that the formulas
exactly correspond to the user input, formula rewriting and
normalization that happened in the underlying logical representation
is hidden from the user.  Finally, when hovering with the cursor on
the inline assertions/assumptions, the corresponding formulas on the
sequent (in case it is needed as a fallback) are highlighted and vice
versa.  In addition, the origins of the formula in the source code are
highlighted.  These highlights enable users to track formulas and
terms and seamlessly switch between the different representations.
\diff{In addition to these presentation features, the prototype also
  lifts the macros that are available in the GUI of \KeY to the source
  code level and makes them available via the context menu, as shown
  in \autoref{fig:screenshot}.  However, the possibility to record
  these high-level interactions as a proof script is still a vision
  and the implementation remains future work.}

While implementing the requirements outlined in \autoref{sec:concept},
we faced several challenges.  We think the solutions and insights
gained are not specific to \KeY, but can also give hints how to
implement our concept in other interactive provers where interactions
are performed on an underlying logical representation different from
the specification
language.%

\subsection{Assigning Initial Origins}
Simplified, the proof obligation in \diff{\KeY's internal logic
  JavaDL} for a functional method contract
$C = (\mathit{pre}, \mathit{post}, \mathit{mod}, \mathit{term})$,
where $C$'s components are \diff{JavaDL's} equivalents to
\lstinline|requires|, \lstinline|ensures|, \lstinline|assignable|, and
\lstinline|measured_by|, for \diff{instance} method \java{m(p$_1$,
  $\ldots$, p$_n$)} is as follows\diff{, where \java{self} is an
  object of the correct type}:
\begin{alignat*}{1}
  &\mathit{pre} \rightarrow\big[\!\!\big[\java{res=self.m(p$_1$, $\ldots$, p$_n$);}\big]\!\!\big] (\mathit{post} \wedge \mathit{frame})
\end{alignat*}
We assign corresponding origin types to the subformulas in
$\mathit{pre}$, $\mathit{post}$, and $\mathit{frame}$ \diff{while
  these terms and formulas are created after parsing the
  specification}.  However, this proof obligation provides only a very
simplified view.  In reality, many technical details have to be taken
into account.  In addition to the original precondition $\mathit{pre}$
of the method, there are many additional formulas added to the premise
of the implication, which are necessary to faithfully model the Java
program in logic.  This includes information about the well-formedness
of the heap ($\dlfunc{wellFormed}(\java{heap})$), that the object
referred to by \diff{\java{self}} is created
($\java{self.created} \keyeq \keybooltrue$), the exact type of
\diff{\java{self}}, and more.  When assigning origin types to these
implicit formulas, it is important to mark them as implicit to later
be able to distinguish them from user inputs.  The particular
challenge for the implementation was that these adaptations had to be
done in Java code in the core of \KeY, since this is where the
generation of proof obligations happens.

\subsection{Transforming Origins Under Sequent Calculus Rules}
After the initial creation of the proof obligation, all subformulas in
the sequent have an origin reference assigned, which already allows
tracking where the formulas come from in the initial proof obligation.
This information in itself can already be helpful for the user, for
example when variable names might be changed during generation of the
proof obligation formula.  However, since the purpose of the origin
information is to allow for tracking in later goals, it is crucial to
know how to transform it under rule applications.

\KeY works deductively with a large set of rewriting and sequent
calculus rules (more than 1\,500).  Most of them are written in a
domain specific language (called ``Taclet language''), and they can be
classified into several categories.  For these categories, we present
a generic scheme that allows to maintain the origins throughout the
proof.  We have to make sure not to accumulate origins too much, a
problem we faced in earlier experiments: After some rule applications,
the origin of even a small subterm could become nearly the whole
method contract, which is not helpful for the user.

The first rule category are pure rewriting rules, which can take a
subterm $s$ of an arbitrary formula on either side of the sequent and
rewrite it into a different term $t$: $s \rightsquigarrow t$.
In this case, we transfer the origin of $s$ to the new term $t$.  The
intuition is that the formula was transformed without dependency to
any additional position in source code.  This can happen for example
for symmetry rules of operators.

The second category is that of conditional rewrite
rules. %
It is similar to pure rewriting rules, but here the rule is only
applicable if an additional condition holds.  This condition is given
as a schematic sequent $\alpha_0, \alpha_1, ... \Longrightarrow$,
where the $\alpha_i$ can be either in antecedent or succedent.
In this case, the new term $t$ gets the combined origins of $t$ and
the $\alpha_i$.
Intuitively, the rewritten term $t$ is the result of combining the
$\alpha_i$ and $s$, and thus gets the union of their origins.  An
example would be when $\alpha$ is an equation in the antecedent that
gets applied in a formula on the subterm $s$.

Apart from this, there are some general schemes we apply to cover
other categories of rules.  First, terms that are newly added by a
rule do not get any origin.  Second, if the \diff{resulting term} ($t$
in the above example) is a subterm of the matched term ($s$), it keeps
its original origin.  This is crucial to prevent from too much
accumulation. %
Third, terms that stem from user input during a rule application (for
example the \texttt{cut} rule) get a corresponding label, but no file
or position information.  Finally, all other terms that are not
touched by a rule keep their origins.

Two rules are essential for verification of programs: rules for loops
and method calls.  Due to their complexity, they are implemented in
Java code within the prover core of \KeY.  Both of them similarly
split the proof into several branches, for instance where the loop
invariant has to be shown to hold before entering the loop, the body
has to be proven to preserve the invariant, and a use case that
continues with the symbolic execution of the remaining program while
assuming the loop invariant.  We assign position information to the
formulas originating from the specification (such as the loop
invariant), and set distinct origin type labels for the generated
terms, which allows tracking them throughout the proof.  A
particularly interesting case is the ``body preserves invariant''
branch, where the invariant formula appears twice and thus has the
same source code position information.  We distinguish them via
different origin types.  In addition, like in the initial proof
obligation, several implicit formulas are added, which we again handle
with specific \enquote{implicit} origin types.

\subsection{Translating Formulas from JavaDL back to JML}
In the optimal case, when a formula has only a single unique origin
\diff{and only refers to a single program state}, its original JML
formula can be shown verbatim in an \diff{assertion or an assumption}
as described in \autoref{sec:concept}.  For cases where this is not
possible, we implemented a translation back to JML.
With that, a particular challenge was that due to how the sequent
calculus works in \KeY, many terms get extracted with the ``pull-out''
rule into an equation with a new Skolem constant.  This increases
efficiency during proof search: If a term occurs multiple times in the
sequent, its simplification only has to be performed once.  However,
this poses a problem for the translation,
\diff{since these Skolem constants only exist on the low-level logical
  representation, but do not have a direct counterpart in JML.  To
  solve this challenge, we had to implement an inlining mechanism that
  reverts the pull-out.}
  
\diff{A caveat is that the translation is not always possible: The
  logical level may contain low-level constructs that cannot be
  represented in JML.  In these cases, the user is presented an
  \lstinline|assert|/\lstinline|assume| with a clear error message,
  and the only way to continue is in the logical representation.
  However, this fall-back should occur rarely (if at all) in practice,
  since it only happens when implicit knowledge \emph{must} be
  rendered but cannot.
  This is only the case if it stems from user input on the low-level logical view, and it is reasonable to continue on this lower level.}

\subsection{Computing Positions for Assertions and Assumptions}
For modeling the heap, \KeY uses the theory of
arrays~\cite{McCarthy63}.  We use explicit terms to represent heap
states. There are functions $\mathtt{select}(h, o, f)$ and
$\mathtt{store}(h, o, f, v)$ to read and write from a heap, where $h$
is a heap, $o$ and object, $f$ a field, and $v$ a value (for
simplicity, we ignore the types here).  An example term might be
\texttt{select(store(heap, self, A::\$f, $5$), self, A::\$f)}, which
with the help of the axiom \texttt{selectOfStore} could be simplified
to 5.  For more details, we refer to~\cite[ch.~2]{ABBHSU2016}.

In general, a sequent of a program proof will contain many of these
terms, corresponding to different chains of read and write statements.
The observation here is that every heap term that occurs in the
sequent corresponds to an exact line in source code.  This is the case
because such terms only get created via symbolic execution of the
program.  If we allowed a user to insert low-level formulas directly
(e.g., via the cut rule), this assumption would break, and the
produced sequent could not be translated any more.  However, in our
interaction concept the user should stay as long as possible (ideally
always) on the code level.

\begin{figure}
\begin{subfigure}{\linewidth}
\centering
\begin{lstlisting}[language=Java, numbers=left]
class C {
  int sum, max;
  void sumAndMax(int[] a) {
    sum = 0;
    max = 0;
    int k = 0;
    while(k < a.length) {
        ...
    }
  }
}
\end{lstlisting}
\caption{Example program with two heap assignments and a loop.}
\label{fig:sum-and-max}
\end{subfigure}
\vspace{0cm}

\begin{subfigure}{\linewidth}
\begin{lstlisting}[language=Java, mathescape=false, escapechar=§]
anon(                            // loop       line 7
  store(                         // assignment line 5
    store(                       // assignment line 4
      heap, self, C::$sum, 0)
    self, C::$max, 0),
  {(self, C::$max)} §$\cup$§ {(self, C::$sum)}
  anon_heap_LOOP)
\end{lstlisting}
\Description{A class C with integer fields sum and max and a method
 called sumAndMax. The method iterates over an array a and computes
 both the sum of all fields and the maximal value of the array
 elements and stores the result in fields sum and max,
 respectively. Below the class is an anonymized heap term, where
 only the locations of sum and max are anonymized.}
\caption{Heap term formalizing all assignments up to line 8 in
  \autoref{fig:sum-and-max} with corresponding line numbers \diff{from
    which the sub-terms originate}.}
\end{subfigure}
\caption{Example of a nested heap term for a program with heap
  assignments and a loop.}
  \label{fig:heap-term}
\end{figure}

An example of a heap term with corresponding line numbers is shown in
\autoref{fig:heap-term}.  The function $anon$ is used to anonymize a
set of locations on the heap (the fields sum and max of \texttt{this}
here).  This is done when concrete code is abstracted by
specifications, that is, for loop invariants and method contracts.

For the translation, we construct a heap map for the sequent.  A very
similar approach is applied for local variables, which are handled in
\KeY by producing multiple copies, such as \texttt{k\_0, k\_1, etc.}
For readability, we omit further details about that here.  With these
maps available, the positioning for assumptions and assertions works
as follows.
On the path through the program corresponding to the current goal,
from the start of the method to the most recently executed line, we
(virtually) insert the assumptions and shift them downwards, until a
statement with a heap change is reached that is not part of the
formula, or the most recently executed statement is reached.  For
instance, a formula containing the outer store term shown in
\autoref{fig:heap-term} and no other heap terms would be positioned
directly in front of the loop, since the start of the loop would be
the next line with a heap attached.
For assertions, this works vice versa: The initial position is the
last line of the method, and they are shifted upwards until they reach
a heap update not contained in the formula or the most recently
executed statement.

Formulas referring to multiple states are inserted into the code as
late as possible, referring to other states with the JML-inspired
operator \lstinline{\old<line>(<term>)}.  This is in contrast to the
concept in \autoref{sec:concept}, where we describe the JML operator
\lstinline{\old(<term>, <label>)}.  However, the latter would require
to introduce many additional labels and sometimes even blocks, which
would modify and clutter the original source code.  Since we only use
the operator for presenting to the user, a line number is more
intuitive and easier to implement correctly.

\subsection{Limitations}
Due to the prototypical state of the implementation, there are
currently some limitations.  First, the implementation only works for
one of the two loop rules; \diff{for the other rule,} a clear error
message is given such that the user can change the settings
accordingly.  Second, goals where the assignable clause has to be
proven cannot be represented in JML at the moment, since the
corresponding formula is not expressible.  Third, when a formula
deviates too much from the original user input or no origin
information is found, a best-effort re-translation has to be
performed, which is currently only implemented partially.  Finally,
the branch corresponding to the framing clause of a method contract or
loop specification can currently not be represented in JML, since the
corresponding JavaDL formula quantifies over fields, which is not
expressible in JML.  As a future workaround, there is a function
$\mathit{assignable}$ that could be used to represent this formula on
JML level, however, this is currently not yet implemented.

\section{User Study}
\label{sec:user-study}
To evaluate the efficacy of the interaction concept and the prototype
implementation on first contact, we conducted a \diff{pilot} user
study with six experts in formal verification, most of them seasoned
users of the \KeY system.  The goal of the study was to find out if
interaction on source code and representing goals inline in source
code can support the user in their verification process. \diff{The
  study is designed to get a first impression of the new view's effect
  on goal state comprehension rather than a representative
  examination.}  At the time the user study was conducted, the
prototype did not yet support proof scripts.

\subsection{Study Design}
\begin{figure*}
  \includegraphics[width=\linewidth]{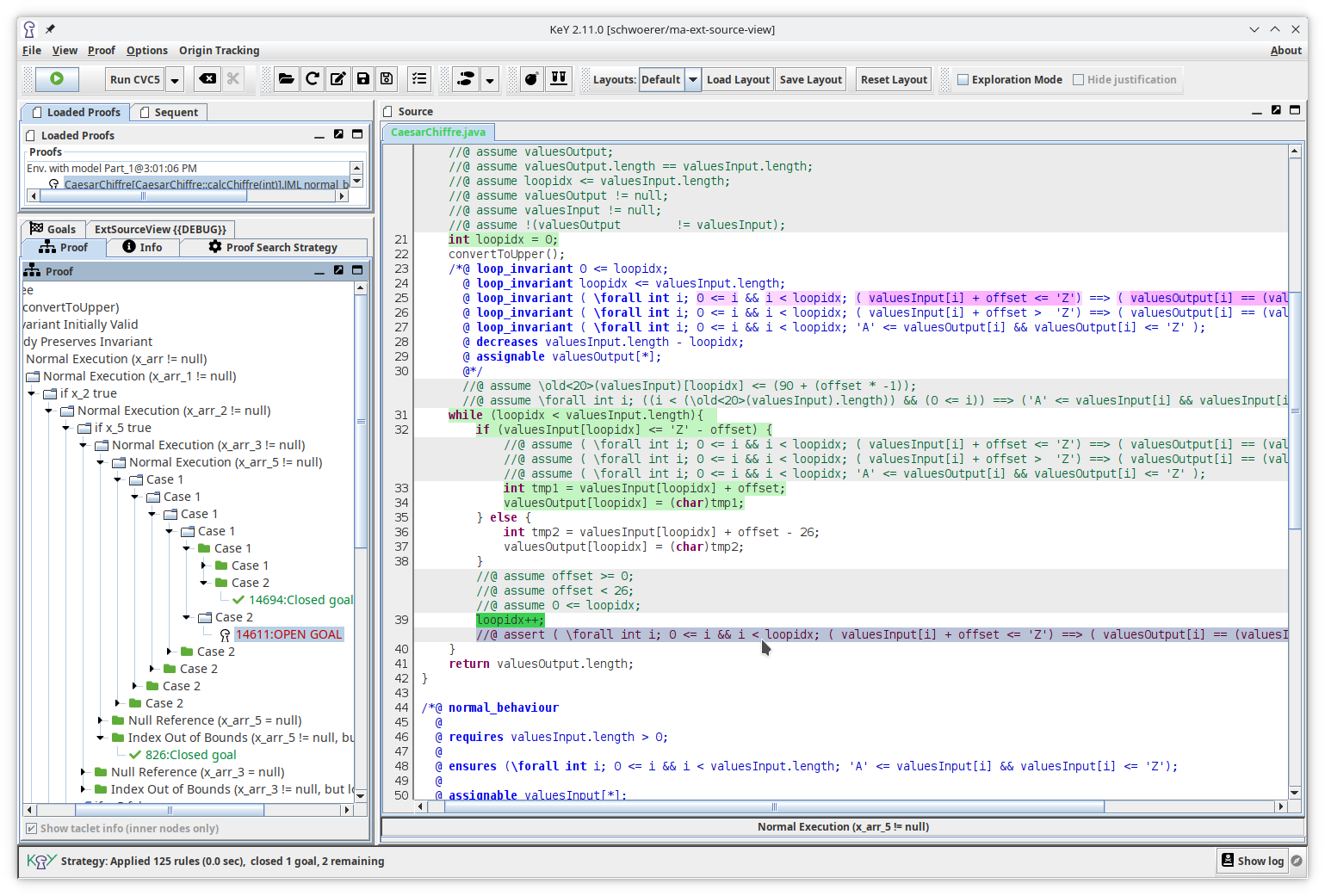}
  \Description{A screenshot of the \KeY GUI taken during the user
    study. It shows the proof tree and the new interaction view.}
  
  \caption{Screenshot of the \KeY GUI showing an open goal state of
    the first task of the user study using the new source code level
    representation.}
  \label{fig:screenshot-eval}
\end{figure*}

Written explanations, source code, a \KeY version, and a questionnaire
were provided to the participants. All participants were sufficiently
familiar with the \KeY system that a general \KeY tutorial was
unnecessary.

The study comprised two verification tasks to be performed on two very
similar variants of a small program with about 40 lines of Java code
and 50 lines of JML specification.  The program implements a Caesar
cipher over a character array and contains two methods, each featuring
a loop and several branching statements.  Both tasks required
correcting a small flaw in either the source code or the
specification; the tasks were designed to be comparable in difficulty.

For the first task, participants were asked to interact with the new
view with inline overlays (the original sequent view was not
permitted) to identify and locate the flaw that prevented
verification.  A screenshot of the GUI with an open goal from this
task is shown in \autoref{fig:screenshot-eval}.  Note that the sequent
view (as normally present in \KeY, compare \autoref{fig:ui}) is
hidden.  Only the source view is shown, where symbolically executed
lines, generated assumptions and assertions, and the origin of the
assertion that is hovered over with the mouse cursor (in magenta) are
shown.

In the second task, the participants performed an
equivalent exercise using the original sequent view; overlays were not
available.  Participants recorded the time they spent on each
task. After completing the experiment, they responded to open‑ended
questions regarding the new interaction concept and its
implementation.

The study design intentionally favored the original sequent view and
interaction in two respects. First, participants were already familiar
with the existing interaction model in \KeY. Second, they encountered
the new features in the first task, so they had to develop their
understanding of the program while using the novel interface. Because
the second task involved essentially the same program, participants
were already familiar with its structure and only had to locate the
second flaw in the sources.

\subsection{Results}

All participants successfully identified the problems in both parts.
Apparently, the new view was sufficiently intuitive: although
intentionally no information was provided explaining the concepts and
features of the new view, all participants were able to intuitively
understand the idea behind the overlaid source annotations, and the
semantics of the new constructs (for instance the \lstinline{\old}
operator with line number), and could explain them correctly in the
questionnaire.

Half of the participants solved task 1 faster than task 2 despite the
missing introduction, the order of the related tasks, and their prior
experience with \KeY.
An overview of the reported times required to finish both tasks can be
found in \autoref{tab:results}.  While the average required time is
slightly larger for task 2 with the old view, the standard deviation
is larger with the new one.
\begin{table}
\begin{tabular}{ c S S }
    \toprule
    Participant     & \parbox[c][][c]{2.5cm}{\centering Task 1\\New Source View\\(min)} & \parbox[c][][c]{2.5cm}{\centering Task 2\\Sequent View\\(min)} \\
    \midrule
    1               & 5.0          & 11.0       \\%
    2               & 16.5         & 11.5       \\%
    3               & 8.0          & 5.0        \\%
    4               & 1.0          & 12.0       \\%
    5               & 16.0         & 8.0        \\%
    6               & 12.5         & 18.5       \\%
    \midrule
    $\mu$           & 9.8          & 11.0       \\%
    $\sigma$        & 5.7          & 4.1        \\%
    \bottomrule
\end{tabular}
\vspace{1em}
\caption{Results of the user study: Reported required times by the individual participants. $\mu$ denotes the average time per task and $\sigma$ is the standard deviation.}
\label{tab:results}
\end{table}

In addition to this quantitative feedback, the study also collected
qualitative feedback in the form of recommendations and remarks.  In
particular, some suggestions were to include the type (for instance,
if a formula stems from an implicit default in JML) in the display.
Also, users particularly liked that technical details needed for the
representation in JavaDL, but not present in JML, are not shown at
all.

Overall, the study shows that the new concept and implementation can
be helpful even for \KeY experts, and that a view closer to the
original representation in the user input can be beneficial for
understanding the sequent.  Since the new interactive view is not
exclusive and can be used in combination with the established forms of
interaction, users can work with what they find more intuitive and
even switch back and forth within a single proof.

\subsection{Validity of the Results}

It should be emphasized that this evaluation was intended to provide
qualitative evidence and to obtain constructive feedback about the
usefulness and understandability of the proposed source code level
interaction concept, not to deliver strong empirical claims about
achievable user performance differences based on the reported times.
Given the small number of participants, the measured task durations
and the observed differences in completion times must be interpreted
cautiously: they illustrate possible benefits rather than constituting
statistically generalizable results. Neither is the sample size
sufficient nor are the participants---being experts in
\KeY---representative of the average verification engineer to draw any
statistical conclusions.  What the study does reliably show is that,
even without prior explanation, participants were able to grasp the
new features added to the tool. \diff{One cannot conclude from this
  pilot study of \KeY experts that a novice to \KeY or to deductive
  verification would benefit from the new view. However, it suggests
  that even experts familiar with \KeY's logical representation and
  internals seem to find the new view helpful. It stands to reason
  that similar benefits can be expected from \KeY beginners, though
  this remains to be tested.}

\section{Related Work} %
\label{sec:related-work}

A number of verification tools are based on the autoactive approach
and, hence, provide interaction patterns similar to our
work. Dafny~\cite{L2010} is a programming language designed to be
easily verified and provides an autoactive verifier. Users typically
stay purely on the source code level and guide the underlying SMT
solver with annotations like \lstinline|assume| and
\lstinline|assert|. Interaction on a lower-level, as in \KeY, is not
possible.

The verification tool Why3~\cite{FP2013} provides autoactive
interaction similar to Dafny. While some interaction on the level of
its programming language WhyML is possible, this is not an end user
language, as it is a functional language with some logical
extensions. Some tools like Creusot~\cite{DJM2022} use it as a
back-end for Rust verification.

On the other hand, VeriFast~\cite{JacobsPiessens08} targets full
program languages like Java, C, and Rust, similar to \KeY. It is
purely an autoactive prover with annotations to guide the prover.

We note that \KeY is the only verification tool targeting an end user
language that provides explicit, fine-grained proof objects,
permitting detailed proof interaction.

JML scripts have a number of precursors and related
approaches. Beckert et al.~\cite{BGU2017} first realized the potential
of extending a verifier such as \KeY with a scripting language,
simplifying manual proof applications. While their scripts offer much
the same expressiveness and power of the JML scripts, the scripts were
external, not part of the specification-level code.

Similar in approach, Grebing et al.\ developed the \emph{Dafny
  Interactive Verifier Environment} (DIVE), which provides a new GUI
for Dafny proofs. This GUI provides access to the source code, a
formula-based logic view, and direct access to proof manipulation in
the form of proof scripts and rule applications~\cite{GKU2020}. While
conceptually close, DIVE differs from our approach in some respects:
DIVE extends the autoactive Dafny system with an explicit logic view
and manual rule applications, whereas \KeY is already based on this
approach and is now equipped with new abstractions, empowering users.

\section{Conclusion}
\label{sec:conclusion}

In this paper, we presented a novel interaction concept integrating
interactive and autoactive verification. It helps to bridge the mental
gap between the source code level, where program and specification are
written, and the logical level, where the proof is conducted when
inspecting and manipulating open proof goals.
We implemented the concept within the deductive Java verifier \KeY and
conducted a small user study which indicates that the novel view of
the program‑verification state can indeed help users better understand
goals and identify and repair flaws in specifications and source code.

A promising idea for future work is to use the new inlined, annotated
view not only for human users but also as input to other tools. For
example, if a goal cannot be proven in \KeY, the view with the
explicit assumptions local to that proof goal could be rendered into a
file which can be loaded into a bounded model checker. Also, this
annotated source file might improve the output of
specification-inference approaches based on large language
models~\cite{BeckertKlamrothPfeifer2024_1000175321}, since it provides
more detailed information about local knowledge.

A second direction is a tighter integration with proofs scripts: The
point-and-click interactions performed on the source view could be
recorded into a script, which should be considerably easier than
persisting low-level sequent interactions in this way.

\diff{Additionally, we aim to expand the pilot study of the new view
  and interaction pattern into a full, reliable, and representative
  user study aimed both at experts and, crucially, \KeY novices. A
  suitable setting would be to test these features in teaching where
  \KeY is already employed and measure the effects of the more
  abstract view on the student's understanding of the proof state---a
  common hurdle for students.}

\begin{acks}
  This work is supported by the \grantsponsor{DFG}{German Research
    Foundation}{https://www.dfg.de/en} under Grants
  No.~\grantnum{DFG}{BE~2334/9-1}, \grantnum{DFG}{BU~2924/3-1},
  \grantnum{DFG}{HA~2617/9-1}, and \grantnum{DFG}{UL~433/3-1}.

  We thank Mike Schwörer for his instrumental work on the new
  interaction view and the user study.
\end{acks}

\bibliographystyle{ACM-Reference-Format}
\bibliography{references}

\end{document}